# Onset of non-diffusive phonon transport in transient thermal grating decay

A. A. Maznev, Jeremy A. Johnson and Keith A. Nelson

Department of Chemistry, Massachusetts Institute of Technology, Cambridge, Massachusetts 02139, USA

The relaxation of a spatially sinusoidal temperature perturbation in a dielectric crystal at a temperature comparable to or higher than the Debye temperature is investigated theoretically. We assume that most phonons contributing to the specific heat have mean free path (MFP) much shorter than the thermal transport distance and can be described by the thermal diffusion model. Low-frequency phonons that may have MFP comparable to or longer than the grating period are described by the Boltzmann transport equation. These low-frequency phonons are assumed to interact with the thermal reservoir of high frequency phonons but not with each other. Within the single mode relaxation time approximation, an analytical expression for the thermal grating relaxation rate is obtained. We show that the contribution of "ballistic" phonons with long MFP to the effective thermal conductivity governing the grating decay is suppressed compared to their contribution to thermal transport at long distances. The reduction in the effective thermal conductivity in Si at room temperature is found to be significant at grating periods as large as 10 microns.

PACS numbers: 66.70.-f, 63.20.-e, 44.10.+i

# I. INTRODUCTION

Thermal transport at small distances is an area of active research.[1-8] In dielectrics and semiconductors heat is carried predominantly by phonons, and thus the relationship between the phonon mean free path (MFP) and the length scale of heat transfer determines whether thermal transport follows the classical thermal diffusion model. The "textbook" room temperature MFP obtained from a simple kinetic model[9] is in the tens of nanometers range even for good heat conductors such as silicon. According to this simplistic view, deviations from the Fourier law at room temperature (RT) would only be expected at distances below ~100 nm.

In reality, describing phonon transport by a single number for the MFP (the so-called gray body approximation) is an oversimplification of the problem. For any length scale there will be phonons of low enough frequency that propagate ballistically rather than diffusively. The question is whether the deviation from the diffusive transport for these low-frequency phonons produces a measurable effect in a given experimental configuration. The issue of the onset of non-diffusive thermal transport is not only of theoretical interest, but also of considerable practical importance in areas such as thermal management of micro- and nano-electronic devices.[2]

The problem of non-diffusive phonon transport at small distances has been subject of extensive theoretical work.[6-8,10,11] However, two obstacles have made it difficult to produce results that could be quantitatively compared to room temperature experimental data. The first factor has been the lack of reliable information on frequency-dependent phonon lifetimes. Until very recently, lifetime information was obtained by fitting thermal conductivity data with multi-parameter Callaway- or Holland-type models.[12,13] Significant progress in this area has now been achieved with the emergence of first-principles calculations of thermal conductivity free of fitting parameters.[8,14-17] As will be shown below, quantitative discrepancies between different models still persist; however, the models invariably point to a large role for low-frequency phonons. For example, according to Henry and Chen,[15] phonons with MFP exceeding 1 μm contribute about 40% to room-temperature thermal conductivity of Si.

Another obstacle has been a gap between theoretical models and experimental configurations. The model favored by theoreticians is that of thermal transport through a slab of material between two black body walls.[6-8,10,11] Reproducing this model experimentally is extremely challenging, since there are no true phonon black body walls and any real interface between two materials involves thermal boundary resistance which by itself presents a long-standing problem in nanoscale thermal transport.[1,18] Thermal measurements across a slab with a fine depth resolution present yet another challenge. Typically, experimental configurations for studying non-diffusive effects in thermal transport at room temperature are not easily subjected to rigorous theoretical analysis, and experimentalists have resorted to simple models based on a modified diffusion theory.[3-5]

An experimental technique for measuring thermal transport in a simple enough geometry to admit a rigorous theoretical treatment does in fact exist under various names including laser-induced transient thermal gratings or impulsive stimulated thermal scattering (ISTS).[19,20] In this method, two short laser pulses of central optical wavelength $\lambda_{opt}$ are crossed in a sample resulting in an interference pattern with period $L = \lambda_{opt}/2\sin(\theta/2)$ defined by the angle $\theta$ between the beams. Absorption of laser light leads to a spatially periodic temperature profile, and the decay of this temperature grating through thermal transport is monitored via diffraction of a probe laser beam. If the grating period is smaller than the absorption depth of the excitation light, thermal transport is nearly one-dimensional and, in a single crystal sample, no interfaces are involved. In addition, a spatially sinusoidal temperature profile facilitates theoretical analysis, as will be shown below. Studying thermal transport on nanometer length scales with transient gratings requires optical wavelengths in the UV or deep UV range, which entails experimental challenges. On the other hand, measurements on ~1 μm scale pose no difficulty and can be done with conventional laser sources in the visible range.[20]

In this paper, non-diffusive phonon transport in thermal grating relaxation at temperatures on the order of or higher than the Debye temperature $\Theta_D$ is studied theoretically. Our focus will be on the onset of non-diffusive transport; the grating period will be assumed to be much larger than MFPs of optical and high frequency acoustic phonons that provide the main contributions to the specific heat. As an example, we will consider grating periods >1 μm in silicon at room temperature (RT).

The main challenge for theoretical analysis of non-diffusive thermal grating decay is the wide range of phonon MFPs, as phonons contributing significantly to thermal conductivity may have their MFP longer than, shorter than, or comparable to the grating peak-null distance $L/2$ over which the heat is transferred, and the transport may vary from purely ballistic to purely diffusive over the phonon spectrum. One approach to handling the wide MFP range is to solve the problem in the "gray-body" approximation, i.e. for a fixed MFP, and then integrate over the phonon spectrum as if phonons of different frequencies contributed to thermal transport independently.[11] This approach may be well warranted when the phonon MFP is dominated by elastic scattering processes such as isotope, defect or grain boundary scattering. However, in single crystal Si and many other materials at room temperature, the dominant scattering mechanism is three-phonon scattering.[21] Thus a more accurate model should include interactions among different parts of the phonon spectrum.

The situation can be simplified significantly if we focus our attention on the onset of non-diffusive transport. Low-frequency phonons with MFP on the order of or longer than ~1 μm in Si at RT may contribute significantly to thermal conductivity but they contribute very little to the specific heat due to their low density of states. Most thermal energy sits in the high-frequency phonon modes with short MFP which can be modeled as a "thermal reservoir" with locally defined temperature and diffusive transport. The concept of separating the phonon spectrum into low- and high-frequency parts goes back to the "two-fluid" model of thermal conductivity.[22] More recently, a "two-channel" model breaking the phonon transport into purely diffusive and purely ballistic components has been applied to the analysis of thermal conductivity in transient thermoreflectance measurements.[3,4] In our model, low-frequency phonons will not be assumed to be purely ballistic. Rather, they will be described by the Boltzmann Transport Equation (BTE) that can handle ballistic, diffusive, and intermediate transport regimes. The high frequency phonons in the "thermal reservoir", on the other hand, will be described by the thermal diffusion equation.

The paper is organized as follows: Sec. II presents the model and main equations; in Sec. III, thermal grating decay is analyzed and the equation for the "correction factor" describing the reduction in the effective thermal conductivity as a function of grating period $L$ is obtained and compared with the results of a "two-channel" model; in Sec. IV an analytical formula for the reduction in the effective thermal conductivity is derived assuming a quadratic frequency dependence of low-frequency phonon lifetimes; in Sec. V a correction due to the Akhiezer mechanism of phonon dissipation at sub-THz frequencies is considered; finally, Sec. VI presents the conclusions of the study.

## II. THE MODEL

We consider a one-dimensional thermal grating created at $t=0$ by short laser pulses crossed in an unbounded material. The details of the laser energy transfer to the lattice are outside the scope of our analysis. In a typical experiment, lased energy is initially absorbed by the electronic sub-system. In many cases, such as intraband carrier relaxation in semiconductors, at least part of the electronic excitation energy is transferred to the lattice on a time scale of ~1ps,[23] hence starting with a spatially sinusoidal phonon population at $t=0$ is not an unrealistic model.

The classical thermal diffusion equation leads to a well known result for the relaxation of the thermal grating with a spatially sinusoidal profile of the temperature perturbation,[19]

$$T = T_0 \exp\left(-\frac{\lambda}{C}q^2 t\right)\cos(qx), \qquad (1)$$

where $q=2\pi/L$ is the "wavenumber" of the thermal grating, $\lambda$ is the thermal conductivity and $C$ is the specific heat per unit volume. In Si at RT, in the range of thermal grating periods $L = 1-10$ μm, the grating decay time according to the diffusion model varies between about 0.3 and 30 ns.

Let us now divide the phonon spectrum into two subsystems: "thermal reservoir" phonons above some frequency $\omega_0$ responsible for most of the specific heat, and low-frequency acoustic phonons below $\omega_0$. The thermal reservoir phonons will have MFP much shorter than the grating period, and thermal transport within the thermal reservoir will be described by the diffusion equation. For the low-frequency part no simplifying assumptions will be made: it will include phonons with MFP greater, on the order of, and shorter than the grating period, and will be described by the BTE. We will see that the exact choice of $\omega_0$ is unimportant and that our results do not depend on $\omega_0$.

The thermal reservoir will be characterized by the temperature $T$ whereas the low-frequency subsystem will be described by the phonon density distribution $n(\omega,\mu)$, where $\omega$ is the frequency and $\mu$ is the unit vector representing the wavevector direction.[24] We will assume that the thermal grating is a small perturbation and hence $T$ and $n$ will denote small deviations from the background equilibrium temperature and phonon density distribution. $n(\omega,\mu)$ is comprised of three sheets corresponding to the three acoustic phonon branches, while all optical modes, as well as high-frequency parts of acoustic branches are included in the thermal reservoir.

For both low frequency and thermal reservoir parts we will adopt the single mode relaxation time approximation (RTA) that was much criticized in the past[21] but is now viewed as adequate for temperatures on the order of or higher than $\Theta_D$. Recent first-principles calculations of the thermal conductivity of Si[14-16] indicated the validity of RTA above ~100 K. Within RTA, lattice thermal conductivity is given by an integral over the phonon spectrum,[25]

$$\lambda = \frac{1}{3} \int_0^{\omega_{max}} c_\omega v \Lambda d\omega \ , \tag{2}$$

where $c_\omega$ is the differential frequency-dependent specific heat per unit volume, $v(\omega)$ is the phonon group velocity and $\Lambda(\omega)$ is the frequency-dependent phonon MFP equal to the product of the velocity $v$ and the relaxation time $\tau(\omega)$. Here and in the following, in order to simplify the notations, summation over phonon branches is implied without an explicit summation sign whenever an integration over phonon frequency is performed. Equation (2) implies that the phonon group velocity and relaxation time are isotropic; however, this assumption is only essential for deriving the final analytical results and the extension of the analysis to a general anisotropic case would be straightforward.

At temperatures on the order of or higher than $\Theta_D$, phonon lifetime is determined mainly by phonon-phonon scattering. We will assume that the low frequency phonons interact with the thermal reservoir of high frequency phonons while the interactions between low-frequency phonons can be neglected. The latter assumption is justified for two reasons: (i) there are fewer low-frequency phonons due do their lower density of states; (ii) the matrix element governing three-phonon scattering is larger for scattering events involving high-frequency phonons.[17,26,27]

Thus the dominant relaxation channel for low-frequency phonons is three-phonon scattering involving one low-frequency phonon and two high-frequency phonons from the thermal reservoir.[22] In other words, relaxation of low-frequency phonons occurs via their absorption or radiation by the thermal reservoir.

Under the above assumptions, the two sub-systems and their interaction in the course of one-dimensional thermal transport are described by the following coupled equations:

$$C\frac{\partial T}{\partial t} = \lambda_r \frac{\partial^2 T}{\partial x^2} - \int d\mu \int_0^{\omega_0} d\omega \frac{n - n_0(T)}{\tau} \hbar\omega$$
$$\frac{\partial n}{\partial t} + v_x \frac{\partial n}{\partial x} = \frac{n_0(T) - n}{\tau(\omega)}$$
(3)

Here $\lambda_r$ is the thermal conductivity of the thermal reservoir given by

$$\lambda_r = \frac{1}{3} \int_{\omega_0}^{\omega_{max}} c_\omega v \Lambda d\omega,$$
(4)

and $n_0(T)$ is the low-frequency phonon distribution in equilibrium with the local temperature $T$. For a small temperature variation,

$$n_0(T) = \frac{1}{4\pi} \frac{c_\omega}{\hbar\omega} T.$$
(5)

The integral in the first line of Eq. (3) represents the rate of the energy loss/gain by the thermal reservoir due to radiation/absorption of low-frequency phonons.

### III. ANALYSIS OF THERMAL GRATING DECAY

Boltzmann transport equation is notoriously difficult to solve even in simple cases such as steady-state transport between two black body walls.[25,28] The difficulty lies in the fact that the phonon distribution function $n$ depends not only on time and coordinate but also on frequency and wavevector direction. The assumption that low-frequency phonon modes interact with the thermal reservoir but not with each other significantly simplifies the problem: we will see that phonons at different frequencies and wavevector directions contribute *additively* to the thermal grating decay. In this section, we present a physically intuitive treatment of the problem, while a more rigorous analysis can be found in the Appendix.

Let us consider the interaction of the thermal reservoir with a sub-group of low-frequency phonons comprising phonons of a specific (longitudinal or transverse) branch at a particular frequency $\omega_i$ and propagating in some specific direction $\boldsymbol{\mu}^{(i)}$ and the opposite direction, $-\boldsymbol{\mu}^{(i)}$. The distribution function $n$ is now comprised of two values, $n_+$ and $n_-$, describing phonon populations moving in the $+\boldsymbol{\mu}^{(i)}$ and $-\boldsymbol{\mu}^{(i)}$ directions, respectively. Equation (3) now takes the form,

$$C\frac{\partial T}{\partial t} = \lambda_r \frac{\partial^2 T}{\partial x^2} + \frac{n_+ - n_0(T)}{\tau}\hbar\omega_i + \frac{n_- - n_0(T)}{\tau}\hbar\omega_i$$
$$\frac{\partial n_+}{\partial t} + v_x \frac{\partial n_+}{\partial x} = \frac{n_0(T) - n_+}{\tau} \quad , \quad (6)$$
$$\frac{\partial n_-}{\partial t} - v_x \frac{\partial n_-}{\partial x} = \frac{n_0(T) - n_-}{\tau}$$

where $v_x = \mu_x^{(i)} v$. Introducing new variables

$$F_S = (n_+ + n_-)\hbar\omega_i$$
$$F_D = (n_+ - n_-)\hbar\omega_i \quad (7)$$

and using Eq. (5), we obtain the following representation of Eq. (6),

$$C\frac{\partial T}{\partial t} = \lambda_r \frac{\partial^2 T}{\partial x^2} - \frac{c_\omega}{2\pi\tau}T + \frac{1}{\tau}F_S$$
$$\frac{\partial F_S}{\partial t} + v_x \frac{\partial F_D}{\partial x} = \frac{c_\omega}{2\pi\tau}T - \frac{1}{\tau}F_S \quad . \quad (8)$$
$$\frac{\partial F_D}{\partial t} + v_x \frac{\partial F_S}{\partial x} = -\frac{1}{\tau}F_D$$

For a thermal grating with the grating wavenumber $q$, we assume a sinusoidal spatial dependence $\exp(iqx)$ of all variables, which reduces Eq. (8) to a set of ordinary differential equations for the amplitudes:

$$\frac{\partial T}{\partial t} = -\frac{1}{C}\left(\lambda_r q^2 + \frac{c_\omega}{2\pi\tau}\right)T + \frac{1}{\tau C}F_S$$

$$\frac{\partial F_S}{\partial t} = \frac{c_\omega}{2\pi\tau}T - \frac{1}{\tau}F_S - iqv_x F_D \qquad . \qquad (9)$$

$$\frac{\partial F_D}{\partial t} = -iqv_x F_S - \frac{1}{\tau}F_D$$

Decay rates $\gamma$ controlling the dynamics of the system are found by equating to zero the determinant:

$$\det\begin{vmatrix} -\frac{1}{C}\left(\lambda_r q^2 + \frac{c_\omega}{2\pi\tau}\right)-\gamma & \frac{1}{\tau C} & 0 \\ \frac{c_\omega}{2\pi\tau} & -\frac{1}{\tau}-\gamma & -iqv_x \\ 0 & -iqv_x & -\frac{1}{\tau}-\gamma \end{vmatrix} = 0 . \qquad (10)$$

Let us assume, initially, that the relaxation time $\tau$ is much shorter than the thermal grating decay time. In this case, we should expect the system dynamics to be comprised of a fast rearrangement on the time scale of $\tau$ followed by a slow decay. If we are only interested in the slow dynamics, we can set $(1/\tau+\gamma)=1/\tau$, which leads to the following result:

$$\gamma = -\frac{\lambda_r q^2}{C} - \frac{c_\omega}{2\pi\tau C}\frac{q^2 v_x^2 \tau^2}{1+q^2 v_x^2 \tau^2} \qquad . \qquad (11)$$

The second term is the contribution of the low-frequency phonons to the decay rate. If the phonon MFP is much smaller than the grating period, $qv_x\tau \ll 1$, the denominator in the second term of Eq. (11) is close to unity. We will see that in this case the correction to the decay rate is in agreement with the diffusion model. The opposite case, $qv_x\tau \gg 1$, corresponds to the ballistic limit.

Now let us consider what happens if $\tau$ is on the order or larger than the thermal grating relaxation time $1/\gamma$. Under the assumptions of Sec. II, the thermal grating decay time must be much longer that the phonon time-of-flight over a distance $L/2\pi$, hence $\gamma \ll qv$. Indeed, the grating decay time on the order of $1/qv$ would imply that the heat is carried at the speed of sound, which is only possible if *all* phonons are ballistic. For example, at $L=10$ μm Eq. (1) yields $\gamma \sim 3\times 10^7$ s$^{-1}$ (we will see that the actual decay rate is smaller) whereas $qv$ exceeds $3\times 10^9$ s$^{-1}$ for both transverse and longitudinal acoustic phonons. Consequently, phonons whose lifetime is on the order of or longer than the grating decay time ought to be ballistic, with $qv\tau \gg 1$. Thus if $\tau \geq 1/\gamma$, then both $\gamma$ and $1/\tau$ in Eq.(10) can be assumed small compared to $qv_x$. In this case, Eq. (10) yields the result

$$\gamma = -\frac{\lambda_r q^2}{C} - \frac{c_\omega}{2\pi\tau C} \quad , \tag{12}$$

which coincides with the ballistic limit of Eq. (11).

According to Eq. (11), low-frequency phonons contribute additively to the thermal grating decay rate. Thus in order to account for the interaction of the thermal reservoir with phonon sub-groups corresponding to different frequencies and angles, we add their respective contributions by integrating over angle and frequency,

$$\begin{aligned}\gamma &= -\frac{\lambda_r q^2}{C} - \int_0^{\omega_0} d\omega \int d\mu \frac{c_\omega}{2\pi\tau C} \frac{q^2 v_x^2 \tau^2}{1+q^2 v_x^2 \tau^2} = \\ &= -\frac{\lambda_r q^2}{C} - \frac{1}{C}\int_0^{\omega_0} d\omega \int_0^{\pi/2} \sin\theta d\theta \frac{c_\omega}{\tau} \frac{q^2 v^2 \tau^2 \cos^2\theta}{1+q^2 v^2 \tau^2 \cos^2\theta}\end{aligned}. \tag{13}$$

A more rigorous derivation of Eq. (13) is presented in the Appendix. The integration over $\theta$ is only from 0 to $\pi/2$ because each phonon subgroup includes phonons traveling in both $+\mu^{(i)}$ and $-\mu^{(i)}$ directions. Assuming that $v$ and $\tau$ are independent of $\theta$ (here is one place where the isotropic approximation is essential), this integration can be performed analytically with the following result:

$$\gamma = -\frac{\lambda_r q^2}{C} - \frac{1}{C} \int_0^{\omega_0} d\omega \frac{c_\omega}{\tau} \left(1 - \frac{\arctan(q\Lambda)}{q\Lambda}\right), \tag{14}$$

where the product $v\tau$ has been replaced by the phonon MFP $\Lambda$. Thus the effective thermal conductivity is given by

$$\lambda_{eff} = \int_{\omega_0}^{\omega_{max}} \frac{1}{3} c_\omega v^2 \tau d\omega + \int_0^{\omega_0} \frac{c_\omega}{\tau q^2} \left(1 - \frac{\arctan(q\Lambda)}{q\Lambda}\right) d\omega. \tag{15}$$

If the MFP is much smaller than the grating period, $q\Lambda \ll 1$, expansion of the arctangent in a Taylor series leads to the following result,

$$1 - \frac{\arctan(q\Lambda)}{q\Lambda} \simeq \frac{1}{3} q^2 \Lambda^2, \tag{16}$$

and the two integrands in Eq. (14) become equal. Therefore the effective conductivity can be represented by a single integral,

$$\lambda_{eff} = \frac{1}{3} \int_0^{\omega_{max}} A c_\omega v \Lambda d\omega, \tag{17}$$

where the correction factor $A$ is given by

$$A(q\Lambda) = \frac{3}{q^2 \Lambda^2} \left(1 - \frac{\arctan(q\Lambda)}{q\Lambda}\right). \tag{18}$$

The dependence of $A$ on $q\Lambda$ is shown by the solid curve in Fig. 1(a). At $q\Lambda \ll 1$ the correction factor becomes unity, hence the contribution of the "diffusive" phonons to the effective thermal conductivity is, as expected, consistent with Eq. (2). At $q\Lambda \gg 1$ the correction factor drops down as $(q\Lambda)^{-2}$ which means that the contribution of low-frequency "ballistic" phonons to $\lambda_{eff}$ is strongly suppressed compared to Eq. (2). The result is a reduction in the effective thermal conductivity due to the reduced contribution of ballistic phonons.

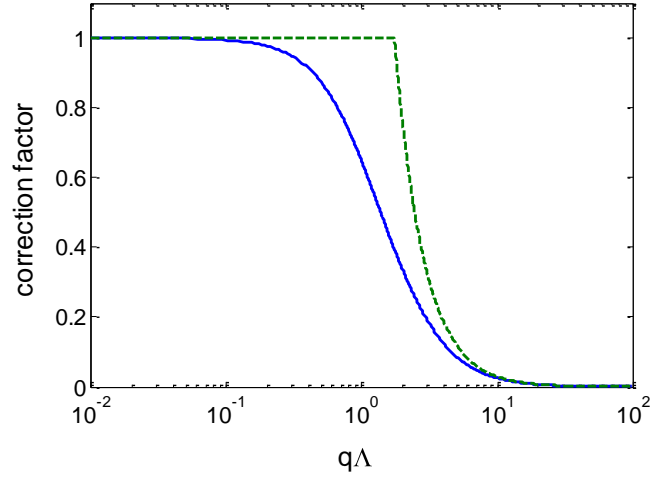

FIG. 1. (Color online) Correction factor *A* vs. the product of the grating wavenumber and phonon mean free path. The solid curve was calculated according to Eq. (18) while the dashed curve was obtained from the "stitching model", Eq. (23).

### A. Ballistic limit and stitching model

Let us consider a "two-channel" model[3,4] where all phonons below $\omega_0$ are assumed to be ballistic. The thermal grating decay rate for this case can be obtained from Eq. (14) assuming ballistic limit $q\Lambda \gg 1$ for the low-frequency phonons. It is instructive to obtain the same result directly from Eq.(3) using intuitive physical considerations. Since ballistic phonons travel over a distance much larger than the grating period, they redistribute energy uniformly over peaks and nulls of the thermal grating. Once a ballistic phonon has been radiated, its energy is lost as far as the thermal grating is concerned. Thus the "grating" component of the ballistic phonon population is at all times zero, and their radiation/absorbtion rate, according to Eq. (3), is given simply by $n_0(T)/\tau$. Making use of Eq. (5), we get the following equation for the temperature of the thermal reservoir,

$$C\frac{\partial T}{\partial t} = \lambda_r \frac{\partial^2 T}{\partial x^2} - \int_0^{\omega_0} d\omega \frac{c_\omega T}{\tau}. \tag{19}$$

Assuming a sinusoidal spatial dependence of the temperature given by exp(*iqx*), we arrive to an ordinary differential equation yielding a relaxation rate

$$\gamma = -\frac{\lambda_r q^2}{C} - \frac{1}{C}\int_0^{\omega_0} \frac{c_\omega}{\tau} d\omega. \tag{20}$$

As expected, exactly the same result can be obtained from Eq. (14) in the ballistic limit $q\Lambda \gg 1$. The effective thermal conductivity is now given by

$$\lambda_{eff} = \int_{\omega_0}^{\omega_{max}} \frac{1}{3} c_\omega v^2 \tau d\omega + \int_0^{\omega_0} \frac{c_\omega}{q^2 \tau} d\omega, \tag{21}$$

with the first term representing the diffusive and the second the ballistic contribution. We can now "stitch" diffusive and ballistic contributions by requiring that at $\omega=\omega_0$ both integrands in Eq. (21) be equal. We get

$$\Lambda(\omega_0) = \frac{\sqrt{3}}{q}. \tag{22}$$

Thus $\omega_0$ is now not arbitrary but determined by the "stitching" condition for each phonon branch. Equation (21) can be represented in the form of Eq. (17) with the correction factor given by

$$A_{stitch}(q\Lambda) = \begin{cases} 1 & \text{if } q\Lambda < \sqrt{3} \\ 3\ q\Lambda^{-2} & \text{if } q\Lambda > \sqrt{3} \end{cases}. \tag{23}$$

Dashed curve in Fig. 1(a) represents the correction factor according to the two-channel stitching model. Naturally it yields the same results as Eq. (18) in the ballistic and diffusive limits; however, the stitching model overestimates the contribution of the intermediate range phonons with $q\Lambda \sim 1$ to the effective thermal conductivity.

## IV. REDUCTION IN THE EFFECTIVE THERMAL CONDUCTIVITY

Equation (17) solves the problem of finding the thermal grating decay rate provided that the phonon density of states, group velocity and relaxation time for all phonon branches are known. For Si at room temperature, these quantities have been calculated in recent first-principles studies.[8,14-17]

To make further progress in the analytical treatment of the problem, we take advantage of the fact that the total thermal conductivity $\lambda$ is normally well known from experiment. Thus instead of calculating $\lambda_{eff}$ according to Eq. (17) it would suffice to calculate the reduction in the effective thermal conductivity due to non-diffusive transport,

$$\Delta\lambda = \lambda - \lambda_{eff} = \frac{1}{3}\int_0^{\omega_{max}} (1-A)c_\omega v^2 \tau d\omega. \qquad (24)$$

The advantage of using Eq. (24) as compared to Eq. (17) is that the factor (1-$A$) is significantly non-zero only at low frequencies where the calculations can be simplified.

To derive an analytical expression for the onset in the reduction of the thermal conductivity, let us assume that the low-frequency phonons are non-dispersive and that the relaxation time depends quadratically on frequency,[15-17]

$$\tau_{l,t} = \frac{1}{a_{l,t}\omega^2}, \qquad (25)$$

where subscripts $l$ and $t$ refer to the longitudinal and transverse branches, respectively. At low frequency and high temperatures, the average thermal energy per phonon mode is $k_B T$, and the differential frequency-dependent specific heat is given by the product of the Boltzmann constant $k_B$ and the low-frequency density of states,

$$c_\omega = \frac{k_B \omega^2}{2\pi^2 v_{l,t}^3}, \qquad (26)$$

The reduction in the effective thermal conductivity is now given by

$$\Delta\lambda = \frac{k_B}{6\pi^2}\sum_{l,t}\frac{1}{a_{l,t}v_{l,t}}\int_0^{\omega_{max}}\left[1-A\left(\frac{qv_{l,t}}{a_{l,t}\omega^2}\right)\right]d\omega, \qquad (27)$$

where the summation over the phonon branches is now incorporated explicitly. Since the integrand is only substantially nonzero at low frequencies, we can extend the integration to $+\infty$, with the following result,

$$\Delta\lambda = \frac{\eta k_B q^{1/2}}{6\pi^2}\sum_{l,t}\frac{1}{a_{l,t}^{3/2}v_{l,t}^{1/2}}$$
$$\eta = \int_0^\infty dx\left[1-3x^4+3x^6\arctan\left(\frac{1}{x^2}\right)\right] \simeq 0.951 \qquad (28)$$

Table I presents values of coefficients $a_{l,t}$ for Si at RT obtained by Henry and Chen[15] and Ward and Broido[16] by fitting first-principles calculation data. Values of the MFP at $\omega/2\pi=1$ THz are also presented. One can see that there is some discrepancy between low-frequency phonon lifetime factors in the two studies, even though both yielded the correct value for the total thermal conductivity. Evidently, an accurate determination of frequency-dependent phonon MFP in the absence of direct experimental measurements is still a challenge for theory.

|  | Henry and Chen | Ward and Broido |
|---|---|---|
| $a_l$ ($10^{-17}$ s) | 2.34 | 1.60 |
| $a_t$ ($10^{-17}$ s) | 6.11 | 2.50 |
| $\Lambda_l$ at 1 THz (μm) | 9 | 13 |
| $\Lambda_t$ at 1 THz (μm) | 2.4 | 5.9 |

Table I. Low-frequency phonon lifetime factors and MFPs at 1 THz according to Henry and Chen[15] and Ward and Broido.[16]

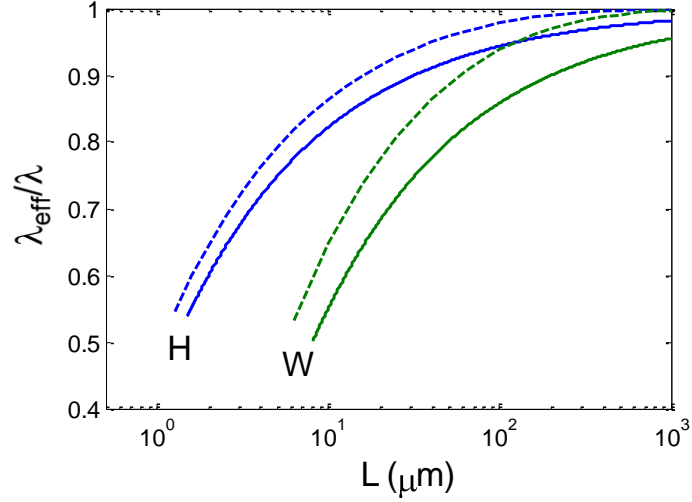

FIG. 2. (Color online) Effective thermal conductivity reduction in Si at RT vs. thermal grating period calculated with the analytical Eq. (28) (solid curves) and corrected for Akhiezer relaxation at low frequencies (dashed curves). Pairs of curves labeled H and W were calculated with phonon lifetime factors taken from Henry and Chen[15] and Ward and Broido,[16] respectively.

The solid curves in Fig. 2 show the reduction in the effective thermal conductivity vs. thermal grating period calculated with Eq. (28) using lifetime factors from Table I and velocity values $v_l=8.4\times10^3$ m/s and $v_t=5.8\times10^3$ m/s. We see that at a grating period as large as 10 μm the model predicts a significant reduction in the effective thermal conductivity, 18% based on Ref. 15 and 45% based on Ref. 16 (since the model is only valid for the *onset* of nondiffusive transport, large reduction figures may be of course inaccurate). It can be shown that for a sinusoidal temperature profile the average distance of the heat transfer is equal to $L/\pi$. Therefore at $L=10$ μm the effective heat transport length is ~3 μm. According to Ref. 15, phonons with MFP over 3 μm contribute 24% to the thermal conductivity of Si at RT. Thus our result is not far from a crude estimate obtained by cutting off the contribution of phonons with MFP larger than the heat transfer length.[3]

## V. EFFECT OF AKHIEZER RELAXATION AT SUB-THZ FREQUENCIES

According to Eq. (28), the reduction in the effective thermal conductivity is inversely proportional to $L^{1/2}$. Thus a measurable effect is expected to be still present at grating periods as large as 1 mm. A similar square root dependence on the heat transfer distance can be seen in the results of numerical calculations of the cross-plane thermal conductivity of thin films[8] which predicted a similarly early onset of the size effect at RT, with ~10% reduction in the thermal conductivity expected for a 100 μm-thick Si film. However, the square root dependence on $L$ results from the assumption that the quadratic dependence of the phonon lifetime on frequency according to Eq. (25) can be extrapolated to arbitrarily low frequencies. This assumption is known to be wrong: in sub-THz range, there is a transition in phonon-phonon interaction from the Landau-Rumer (three-phonon scattering) to the Akhiezer relaxation mechanism[27], and the frequency dependence of the phonon lifetime strongly deviates from the quadratic one.[29]

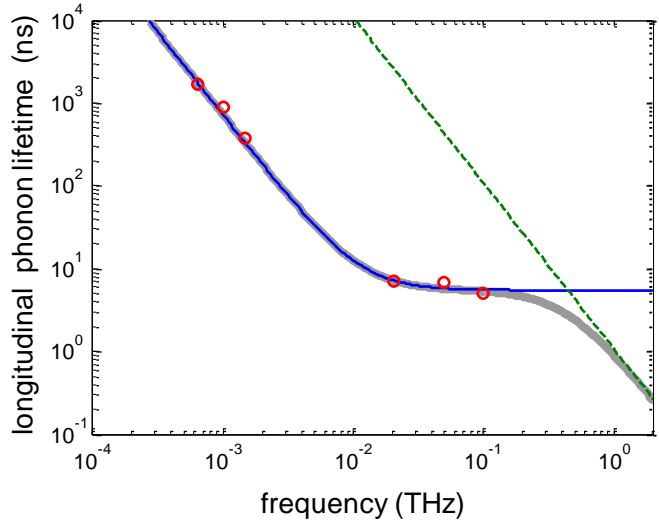

FIG. 3. (Color online) Frequency dependence of the longitudinal phonon lifetime in Si at RT: experimental data from Ref. 29 (symbols) fitted by the Akhiezer relaxation model (thin solid curve); quadratic dependence from the thermal conductivity model[15] (dashed line); total relaxation time according to Eq. (30) (thick gray curve).

Figure 3 shows experimental data on longitudinal phonon lifetime compiled by Daly et al.[29] alongside the quadratic dependence according to Henry and Chen.[15] At ultrasonic frequencies (~1 GHz and below), the dependence is also quadratic;[27] however, there is a disconnect of about three orders of magnitude between ultrasonic data and extrapolated thermal conductivity models. There must be a transition between the two regimes and, indeed, measurements at frequencies 20-100 GHz[29,30] indicated that this transition occurs in the sub-THz range. Currently we do not have any experimental data above 100 GHz tracing the transition into the THz range. For the purposes of the current study we will construct a simple model describing the transition between the two regimes. At frequencies below 100 GHz, the phonon lifetime is determined by the Akhiezer mechanism and can be modeled by a relaxation-type equation,[27,29]

$$\tau_A = \tau_{inf}\left(1 + \frac{1}{\tau_{th}^2 \omega^2}\right), \qquad (29)$$

where the relaxation time $\tau_{th}$ is on the order of the characteristic lifetime of dominant thermal phonons. With $\tau_{inf} = 5.5$ ns and $\tau_{th} = 14$ ps we can reasonably fit the experimental data, as shown in Fig. 3. We will now combine Eq. (29) and Eq. (25) using a Matthiessen-type rule to estimate the total longitudinal phonon lifetime $\tau_l$,

$$\frac{1}{\tau_l} = \frac{1}{\tau_A} + \frac{1}{\tau_{tc}}, \qquad (30)$$

where $\tau_{tc} = 1/a_l \omega^2$. The resulting dependence shown by the gray curve in Fig. 3 provides a smooth transition between the Akhiezer relaxation governing low-frequency behavior and three-phonon scattering dominant at high frequencies.

For transverse phonons, only low-frequency lifetime data at ~1 GHz and below are available. To estimate transverse phonon lifetime within the Akhiezer relaxation model, we adjust the value of $\tau_{inf}$ in Eq. (29) to fit the experimental transverse lifetime data while leaving the relaxation time $\tau_{th}$ unchanged. Since the low-frequency transverse lifetime is about five times longer that the longitudinal lifetime,[31] we simply increase $\tau_{inf}$ five-fold to 27.5 ns. An equation similar to Eq. (30) is then used to estimate the total transverse lifetime.

Now that we have a model for low frequency behavior of phonon lifetimes $\tau_{l,t}$, we can calculate the effective thermal conductivity reduction by an equation similar to Eq. (27), but without making use of Eq. (25),

$$\Delta\lambda = \frac{k_B}{6\pi^2}\sum_{l,t}\frac{1}{v_{l,t}}\int_0^\infty \left[1 - A(qv_{l,t}\tau_{l,t})\right]\omega^2 \tau_{l,t} d\omega. \tag{31}$$

The results of this calculation are shown by dashed curves in Fig. 2. We see that accounting for the Akhiezer mechanism results in a small correction with respect to the total thermal conductivity; however, now no size effect is seen at $L = 1$ mm, due to the reduction in the phonon MFP at sub-THz frequencies.

## VI. CONCLUSIONS

We used to think of heat transport at small distances in terms of the transition from diffusive to ballistic transport depending on the relationship between an average MFP and the distance scale.[10] For phonon-mediated thermal transport, this picture is inadequate due to the wide range of phonon MFPs involved. Rather, one needs to simultaneously consider phonons with MFPs smaller, larger and on the order of the distance scale. To describe the onset of non-diffusive transport, we have proposed a model in which high-frequency phonons responsible for most of the specific heat are described by the thermal diffusion equation whereas the low-frequency phonons are described by the Boltzmann transport equation. The coupled equations of the model have been solved analytically for the transient grating configuration with a sinusoidal temperature distribution. We have shown that the contribution of the "ballistic" phonons to thermal transport in this configuration is reduced compared to the predictions of the thermal diffusion model, leading to a reduction in the effective thermal conductivity. Thus the onset of the size effect is expected to manifest itself in a reduction in the diffusive transport rate rather than a transition to ballistic transport. Assuming a quadratic dependence of the low-frequency phonon lifetime on frequency, we obtained an analytical formula for the reduction in the effective thermal conductivity, which yields an $L^{-1/2}$ dependence on the grating period. In silicon at RT, the thermal conductivity reduction is expected to be significant for grating periods as large as 10 μm. For very large periods this result needs to be corrected by accounting for a reduced phonon lifetime in the sub-THz range due to the Akhiezer dissipation mechanism. However, at $L \sim 10$ μm the correction is small and our analytical formula is expected to be adequate. The analysis presented here lays the groundwork for studying size effects in phonon transport by the transient thermal grating method.

## ACKNOWLEDGMENTS

The authors greatly appreciate the critical reading of the manuscript and illuminating suggestions by Andreas Mayer, as well as stimulating discussions with Gang Chen, David Broido, Arthur Every, and the participants of S³TEC spectroscopy group meetings. This work was supported as part of the S³TEC Energy Frontier Research Center funded by the U.S. Department of Energy, Office of Science, Office of Basic Energy Sciences under Award Number DE-SC0001299.

**APPENDIX**

Let us consider the interaction of the thermal reservoir with all phonon subgroups. Then Eq. (9) will turn into an infinite set of equations. Assuming an exponential temporal dependence $\exp(\gamma t)$ of all variables, we get

$$\left[-\frac{1}{C}\left(\lambda_r q^2 + \sum_i \frac{c_\omega^{(i)}}{2\pi\tau_i}\right) - \gamma\right] T + \sum_i \frac{1}{\tau_i C} F_S^{(i)} = 0$$

$$\cdots\cdots\cdots\cdots\cdots\cdots\cdots\cdots\cdots\cdots\cdots$$

$$\frac{c_\omega^{(i)}}{2\pi\tau} T - \left(\gamma + \frac{1}{\tau_i}\right) F_S^{(i)} - iqv_x^{(i)} F_D^{(i)} = 0 \qquad , \qquad (A1)$$

$$-iqv_x^{(i)} F_S^{(i)} - \left(\gamma + \frac{1}{\tau_i}\right) F_D^{(i)} = 0$$

$$\cdots\cdots\cdots\cdots\cdots\cdots\cdots\cdots\cdots\cdots\cdots$$

where the summation over $i$ is equivalent to an integration over solid angle and frequency and the bottom pair of equations is repeated an infinite number of times. The coefficient matrix of this system of linear equations is block-diagonal, except for the first row and the first column, and hence easy to diagonalize. This block-diagonal form is the consequence of the assumption that low-frequency modes do not interact with each other. It is straightforward to exclude each pair of variables $F_D^{(i)}$ and $F_S^{(i)}$ using the respective pair of equations, which leads to the following equation for decay rates:

$$\gamma = -\frac{1}{C}\lambda_r q^2 - \sum_i \frac{c_\omega^{(i)}}{2\pi C \tau_i} \frac{\gamma^2 + \frac{\gamma}{\tau_i} + (qv_x^{(i)})^2}{\left(\gamma + \frac{1}{\tau_i}\right)^2 + (qv_x^{(i)})^2} \qquad (A2)$$

We are only interested in the slow dynamics with $\gamma \ll qv$, as explained in Sec. III. Under this assumption Eq. (A2) takes the form

$$\gamma = -\frac{1}{C}\lambda_r q^2 - \sum_i \frac{c_\omega^{(i)}}{2\pi C \tau_i} \frac{\frac{\gamma}{\tau_i} + (qv_x^{(i)})^2}{\frac{1}{\tau_i^2} + (qv_x^{(i)})^2}, \tag{A3}$$

which leads to the following result,

$$\gamma\left(C + \sum_i \frac{c_\omega^{(i)}}{2\pi}\frac{1}{1+(qv_x^{(i)}\tau_i)^2}\right) = -\lambda_r q^2 - \sum_i \frac{c_\omega^{(i)}}{2\pi\tau_i}\frac{(qv_x^{(i)}\tau_i)^2}{1+(qv_x^{(i)}\tau_i)^2}. \tag{A4}$$

The sum $\sum_i c_\omega^{(i)}/(2\pi)$ is the total specific heat of the low-frequency phonons. Since this is assumed to be negligible compared to the specific heat of the thermal reservoir, the correction to the specific heat on the left-hand side of Eq. (A4) can be neglected, yielding the final result,

$$\gamma = -\frac{\lambda_r q^2}{C} - \frac{1}{C}\sum_i \frac{c_\omega^{(i)}}{2\pi\tau_i}\frac{(qv_x^{(i)}\tau_i)^2}{1+(qv_x^{(i)}\tau_i)^2} \tag{A5}$$

Replacing the summation by the integration over frequency and angle results in Eq. (13).

24. In order to simplify the notation we use the phonon density rather than the more commonly used occupation number. Our phonon density *n* is defined as the occupation number times the density of states.

25. G. Chen, *Nanoscale Energy Transport and Conversion* (Oxford University Press, New York, 2005).

26. C. Herring, Phys. Rev. **95**, 954 (1954).

27. H. J. Maris, in *Physical Acoustics*, edited by W. P. Mason and R. N. Thurston (Academic, New York, 1971), Vol. 8, p. 279.

28. M. F. Modest, *Radiative Heat Transfer* (Academic, Amsterdam, 2003).

29. B. C. Daly, K. Kang, Y. Wang, and D. G. Cahill, Phys. Rev. B **80**, 174112 (2009).

30. J.-Y. Duquesne and B. Perrin, Phys. Rev. B **68**, 134205 (2003).

31. B. G. Helme and P. J. King, phys. stat. sol. (a) **45**, K33 (1978).